\documentclass[rmp,aps,twocolumn]{revtex4}
\usepackage{graphicx}
\usepackage{amsmath,amssymb}

\begin{document}
\title{Instabilities in the transient response of muscle}
\author{Andrej Vilfan}
\altaffiliation[Present address: ]{J. Stefan Institute, Jamova 39, 1000
  Ljubljana, Slovenia}
\author{Thomas Duke}
\affiliation{Cavendish Laboratory, Madingley Road, Cambridge CB3 0HE, UK}
\email{av242@phy.cam.ac.uk}
\date{20.~March 2003}

\begin{abstract}
  We investigate the isometric transient response of muscle using a
  quantitative stochastic model of the actomyosin cycle based on the swinging
  lever-arm hypothesis. We first consider a single pair of filaments, and show
  that when values of parameters such as the lever-arm displacement and the
  crossbridge elasticity are chosen to provide effective energy transduction,
  the $T_2$ curve (the tension recovered immediately after a step displacement)
  displays a region of negative slope. If filament compliance and the discrete
  nature of the binding sites are taken into account, the negative slope is
  diminished, but not eliminated. This implies that there is an instability in
  the dynamics of individual half-sarcomeres. However, when the symmetric
  nature of whole sarcomeres is taken into account, filament rearrangement
  becomes important during the transient: as tension is recovered, some
  half-sarcomeres lengthen while others shorten. This leads to a flat $T_2$
  curve, as observed experimentally. In addition, we investigate the isotonic
  transient response and show that for a range of parameter values the model
  displays damped oscillations, as recently observed in experiments on single
  muscle fibers. We conclude that it is essential to consider the collective
  dynamics of many sarcomeres, rather than the dynamics of a single pair of
  filaments, when interpreting the transient response of muscle.
\end{abstract}

\keywords{Muscle, Transient response, Myosin, Sarcomere, Biological models,
  Nonlinear dynamics}

\maketitle

\section{Introduction}

The transient response of muscle to a sudden adjustment of its extension, or to
an abrupt change in load, has been one of the most important sources of
information about the mechanism of contraction for over three decades. Ever
since the pioneering work of \citet{huxley71}, experimental data on transients
has informed theoretical models of the interaction between myosin and actin
\citep{huxley71,hill74,eisenberg80,chen93,Huxley2000}, providing a more
detailed picture than could be obtained from the force-velocity relation
\citep{hill38,huxley57} alone. The reason is that the actomyosin interaction
involves several processes which occur on different time scales, and these
individual components can be resolved during the transient response
\citep{huxley71,ford77,Ford.Simmons1985,Ford.Simmons1986,Brenner1991,brenner95}.
The quickest process is the elastic deformation of the myosin crossbridges that
link the thick and thin filaments. Rapid transitions between two or more
different bound states of the myosin molecule are thought to be the next
fastest events, while detachment and reattachment of myosin heads occur on a
slower time scale.

In an experiment to determine the isometric transient response, a muscle fiber
is held at both ends to prevent it from contracting.  The muscle is then
suddenly shortened (or stretched) by a fixed amount, and the tension $T$ that
it generates is measured. Immediately after the imposed change of length, the
tension shifts from the isometric value $T_0$ to a new value, which is termed
$T_1$. But shortly afterwards (typically within $2\,{\rm ms}$), the tension
adjusts to a new value, termed $T_2$.  Subsequently, it gradually reverts to
the original isometric value $T_0$, and the entire transient response is
usually completed in a fraction of a second.  It is generally accepted that the
initial response $T_1$ corresponds to the mechanical deformation of
cross-bridges and provides a direct measure of their elasticity
\citep{huxley71}. The interpretation of $T_2$ is rather more controversial.  It
is often attributed to force generation by the working stroke of bound myosin
molecules
\citep{huxley71,hill74,eisenberg80,Huxley.Tideswell1996,brenner95,duke99,duke2000}
and this interpretation has recently gained support from X-ray interference
techniques applied to shortening fibers
\citep{Irving.Lombardi2000,Piazzesi.Irving2002}. But alternative models suggest
that the force regeneration might be due, in part, to the rapid binding of new
myosin heads to the thin filament \citep{Brenner1991,Howard_book}, or that it
might involve the activation of the second myosin head
\citep{Huxley.Tideswell1997}.

In this article, we wish to address a fundamental problem connected with the
interpretation of force transients.  The present theories are all based on the
consideration of a single pair of filaments, i.e.~one filament containing
myosin molecules, interacting with one actin filament. The dynamics of this
filament pair is generalized to that of a whole muscle fiber by assuming that
all filament pairs in a fiber behave in exactly the same way.  This assumption
is certainly justified as long as there are no static or dynamic instabilities
in the system.  However, the possibility of such instabilities has been known
for a long time \citep{hill74}. Moreover, a stochastic model of the actomyosin
cycle, based on the swinging lever-arm hypothesis, has shown that instabilities
do arise when values of parameters such as the lever-arm displacement and the
crossbridge elasticity are chosen to provide effective energy transduction
\citep{duke99,duke2000}. Such instabilities would give rise to a region of
negative slope in the $T_2$ curve of a single filament pair. Several reasons
have been advanced for the absence of any negative slope in the experimentally
determined $T_2$ curve. \citet{huxley71} argued that the power stroke is
sub-divided into several small steps, and fixed the step size so that the $T_2$
curve had zero slope for limitingly small changes of length. In the model
proposed by \citet{eisenberg80}, the flatness of the $T_2$ curve was explained
by a broad distribution of cross-bridge strain after attachment, combined with
a specific strain-dependence of the transition rates to ensure the proper
occupancies of the two bound states. A further explanation involved the
compliance of the filaments and the distribution of binding sites on the thin
filament in addition to a sub-divided power stroke
\citep{Huxley.Tideswell1996}.  \citet{duke99} has suggested that the flat $T_2$
curve of a muscle fiber can arise despite an instability in the dynamics of a
single pair of filaments, owing to the symmetry of a sarcomere.  We investigate
this possibility further in this article.

\section{Methods}

Simulation of the stochastic evolution of the system was performed using the
Gillespie kinetic Monte Carlo algorithm, which works as follows.  In each
simulation step the rates of all possible transitions are calculated.  The time
until the next event is chosen as a random number with an exponential
distribution and the expectation value given by the inverse of the sum of all
rates. The event itself is chosen randomly with a statistical weight
proportional to its rate.  In the situations with stiff (non-compliant)
filaments and continuous binding sites the transition rates can be factorized
into factors that depend only on the backbone position (which are the same for
all motors in a group) and factors that only depend on the binding position of
a motor (which does not change with time unless that motor undergoes a
transition).  This allowed us to use a very efficient ($n \log n$) algorithm
based on binary trees.

We assumed complete mechanical relaxation of the system in each step, i.e. the
strain of all elastic elements is equilibrated before the next transition takes
place.  The structure involving sarcomeres, filaments and myosin heads was
described as a circuit of elements with given resting lengths and compliances.
The strain of every cross-bridge was calculated, given the constraint of fixed
total length of the system (isometric conditions), or of fixed force acting on
the ends of the system (isotonic conditions).

The $T_2$ transients were always determined $2\,{\rm ms}$ after the
stretch/release.  All other parameter values are summarized in Table
\ref{tab:parameters}.

\begin{table*}
  \caption{Model parameters}
  \label{tab:parameters}
  \begin{flushleft}
  \begin{tabular}{lll}
    \hline
    Cross-bridge spring constant & $K$ & $2.5\,(1.5\footnote{ The numbers in
brackets show an alternative parameter set,
  corresponding to the oscillating case }
)\,{\rm pN/nm}$ \\
    Power-stroke & $d$ & $8\,{\rm nm}$ \\
    & $\delta$ & $0.328\,{\rm nm}$ \\
    Free energy gain& $\Delta G_{\rm stroke}$ & $60\,{\rm pNnm}$ \\
    Transition rates& $k_{\rm bind}$ & $40\,{\rm s^{-1}}$\\
    & $k_{\rm unbind}$ & $2\,{\rm s^{-1}}$\\
    & $k_{1\to 2}$ & $1000\,{\rm s^{-1}}$\\
    & $k_{-ADP}$ & $80\,{\rm s^{-1}}$\\
    Thermal energy& $k_B T$ & $4.14\,{\rm pNnm}$\\
    Dimensionless parameters& $\epsilon_0={Kd^2 }/{2 k_BT}$ & 
$19.3\,(11.6\footnotemark[1])$ \\
    &$\epsilon_1 = K d^2/2\Delta G_{\rm stroke}$ & $1.33\,(0.8\footnotemark[1])$
\\
    &$\epsilon_2 = K d \delta/k_B T$ & $1.6\,(0.95\footnotemark[1])$ \\
    &$\gamma_1 = k_{\rm -ADP}/k_{\rm bind}$ & $2$\\
    &$\gamma_2 = k_{\rm unbind}/k_{\rm bind}$ & $0.05$\\
    Spacing between binding sites & $c$ & $5.5\,{\rm nm}$\\
    Effective lateral crossbridge bending stiffness& $K_A$ & $ 15 \,{\rm pNnm}$\\
    Actin elasticity& $\gamma$ & $44000\,{\rm pN}$\\
    \hline
  \end{tabular}
\end{flushleft}
\end{table*}

\section{Motors acting between a single pair of filaments}
\subsection{Isometric transient in a swinging lever-arm model}

\begin{figure}
   \begin{center}
     \includegraphics{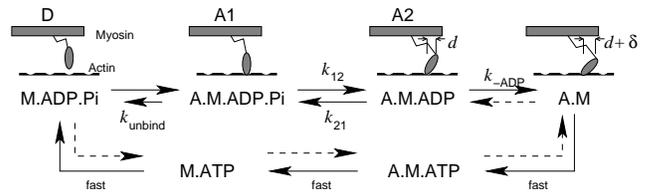}
  \end{center}
  \caption{Reaction scheme for a myosin head.  The dashed lines
    represent transitions which are considered sufficiently slow to be
    neglected in the calculation.}
  \label{fig:reaction_scheme}
\end{figure}

We model the chemical cycle of a myosin head as described in Refs.\ 
\citep{duke99,duke2000} and shown in Fig.~\ref{fig:reaction_scheme}.  A
molecule can exist in either a detached state (D) or in one of the attached
states, A1 (with ADP$\cdot$Pi) or A2 (with ADP).  The state A1 is also referred
to as the weak-binding state \citep{Brenner1991}. The chemical transition
between states A1 and A2 is concomittant with a conformational change of the
molecule, in which the lever arm moves through displacement $d$. From state A2
the head undergoes another conformational change with a lever-arm shift
$\delta$, associated with ADP release.  In the ATP-rich physiological
environment, this step is quickly followed by detachment of the head and we
therefore skip the transitional state (A3) in our model. The attachment and
detachment rates, which are summarized in Table \ref{tab:parameters}, determine
the shape of the force-velocity relation, but have little influence on the
transient response to length steps, which mainly depend on the power-stroke
displacement $d$ and the elastic constant $K$ of the myosin cross-bridge. We
assume that the transitions A1$\leftrightarrow$A2 take place on a faster time
scale than the detachment and reattachment of heads.  Therefore, once a head
has bound with strain $\xi$ in the state A1, its state can be described as a
statistical ensemble of the states A1 and A2 with probabilities given by the
Boltzmann factors
\begin{equation}
\label{eq_p2}
P_2(\xi)=\left( 1+ \exp\frac{-\Delta G_{\rm stroke}+\frac12 Kd^2+Kd \xi}{k_B T}
\right)^{-1}
\quad ,
\end{equation}
where $\Delta G_{\rm stroke}$ is the free energy change associated with
phosphate release.

Now consider a filament containing $N$ myosin motors interacting with a actin
filament which is held to prevent it from sliding. If $N_A$ heads are bound to
the actin with a distribution of strains $\Phi(\xi)$ (normalized to 1), then
the isometric force that they generate is
\begin{equation}
\label{eq_f}
T_0=N_A \int \Phi(\xi) K (\xi + P_2(\xi)d) d\xi \quad .
\end{equation}
If a stretch $\Delta x$ is suddenly applied to the pair of filaments, the force
will first change to
\begin{equation}
T_1(\Delta x)=T_0+N_A K \Delta x \quad ,
\end{equation}
as the cross-bridges are deformed. In the next instance, transitions between
states A1 and A2 occur and the probability distribution re-equilibrates to
$P_2(\xi+\Delta x)$. As a result the force adjusts to
\begin{equation}
  \label{eq_t2}
  T_2(\Delta x)=N_A \int \Phi(\xi) K 
  (\xi +\Delta x + P_2(\xi+\Delta x)d) d\xi \;.
\end{equation}

As a first approximation, we can neglect the distribution of strains and assume
$\xi=0$ for all myosin heads, which allows us to calculate the $T_2$ curves
analytically.  Their shape depends on two dimensionless parameters,
\begin{align}
  \epsilon_0&=\frac{Kd^2 }{2 k_BT} \quad ,
  \label{eq_epsilon0}\\
  \epsilon_1&=\frac {K d^2}{2\Delta G_{\rm stroke}} \quad ,
  \label{eq_epsilon1}
\end{align}
 
which provide a measure of the energy stored in the elastic element when the
head performs a power stroke (A1$\to$A2). The parmeter $\epsilon_0$ measures
this energy relative to the thermal energy, while $\epsilon_1$ compares it to
the magnitude of the chemical free-energy change $\Delta G_{\rm stroke}$ that
accompanies the power-stroke transition. Note that if $\epsilon_1<1$, the power
stroke can occur immediately after the myosin head binds to the thin filament;
but if $\epsilon_1>1$, the conformational change is energetically inhibited
initially, and it is only likely to occur once the thin filament has been moved
forward by the action of other motors. On grounds of efficiency of energy
transduction, we expect the value of $\epsilon_1$ to be as high as possible,
while ensuring that there is sufficient chemical energy to drive the power
stroke, i.e. $\epsilon_1\approx 1$ \citep{duke99}. The power-stroke
displacement $d$ has been measured directly in a number of single-molecule
experiments, and the values obtained range between $5$ and $10\,{\rm nm}$
(reviewed by \citeauthor{Tyska.Warshaw2002}, \citeyear{Tyska.Warshaw2002}).
Data on the crossbridge elasticity $K$ are less reliable, because there are
inevitably other sources of compliance in the system \citep{veigl98}.  But
indirect evidence is provided by the energetic efficiency of muscle, which
peaks at about $50\,\%$
\citep{Kushmerick.Davies1969,Barclay1998,Piazzesi.Lombardi2002}.  Thus a lower
estimate for the energy stored in the elastic element following the power
stroke is one half of the free-energy change accompanying the hydrolysis of an
ATP molecule $\Delta G_{\rm ATP} \approx 20\,{\rm k_BT}$, which leads to
$\epsilon_0\gtrsim 10$.  Taking into account that not all energy stored in the
spring can be converted to mechanical work one obtains a better estimate
$\epsilon_0 \approx 20$ \citep{duke99} (which is consistent with
$\epsilon_1\approx 1$ if most of the energy of hydolysis is used to power the
stroke $\Delta G_{\rm stroke} \approx \Delta G_{\rm ATP}$). This corresponds to
an elastic constant $K=2.5\,{\rm pN/nm}$ if a power-stroke distance of
$d=8\,{\rm nm}$ is assumed.

The form of the $T_2$ curve depends critically on the values of these
dimensionless parameters.
 
If $\epsilon_0>2$, an interval with a negative slope (a hysteresis) exists
\citep{hill74}.  This means that under isotonic (constant load) conditions the
system can be \textit{bistable}.  The location of the interval of negative
slope depends on the value of $\epsilon_1$.  If the hysteresis spans the
origin, then the state with $\Delta x=0$ will be unstable under isotonic
conditions.  Putting $\left. \frac{d T_2}{d (\Delta x)} \right|_{\Delta x=0} <
0$ we see that this occurs if
\begin{multline}
  \left( 1- \frac 1 {\epsilon_0} \ln \left( \epsilon_0-1
      -\sqrt{\epsilon_0\left( \epsilon_0-2 \right) } \right)\right)^{-1} <
  \epsilon_1 \\ < \left( 1- \frac 1 {\epsilon_0} \ln \left( \epsilon_0-1
      +\sqrt{\epsilon_0\left( \epsilon_0-2 \right) } \right)\right)^{-1}
  \label{eq:neg_slope}
\end{multline}
With a value of $\epsilon_1\approx 1$ the $T_2$ curve always has a negative
slope around the stationary point.  For example, with $\epsilon_0=20$,
Eq.~(\ref{eq:neg_slope}) yields $0.85<\epsilon_1<1.22$.  However, it should be
bourne in mind that this calculation does not take into account the
distribution of strains on cross-bridges within the ensemble, and so the actual
range might deviate slightly from this estimate.

\begin{figure}
   \begin{center}
     \includegraphics{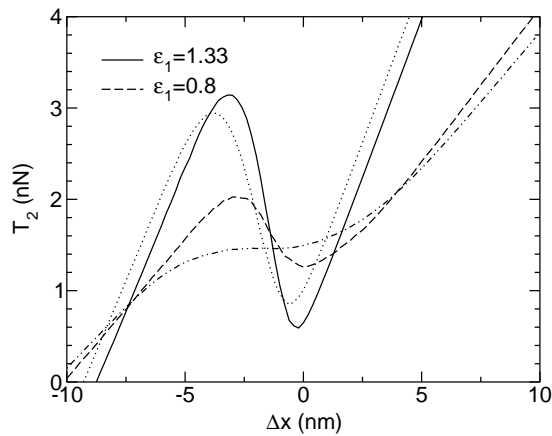}
  \end{center}
  \caption{The $T_2$ transient response to a step-wise stretch for a group of
    300 motors in the isometric state for two different values of the parameter
    $\epsilon_1$.  The dotted and the dashed-dotted lines shows the response of
    a group of motors which are pulling against an elastic element instead of
    being completely isometric before the length step (although the length step
    is then imposed on the motors alone, without the elastic element).  Other
    parameters were: $\Delta G_{\rm stroke}/k_BT=15$, $\epsilon_0=19.3$,
    $\epsilon_2=Kd \delta/k_B T=1.6$ (when $\epsilon_1=1.33$) or
    $\epsilon_0=11.6$, $\epsilon_2=Kd \delta/k_B T=0.95$ (when
    $\epsilon_1=0.8$), $\gamma_1=2$ and $\gamma_2=0.05$.}
  \label{fig_t2}
\end{figure}

An example of the $T_2$ transient for a group of myosin heads between two
firmly clamped filaments is shown in Fig.\ \ref{fig_t2}.

\subsection{Group of stiffly coupled motors under near-isometric conditions}
\label{sec_osci}

\begin{figure*}
  \begin{center}
    \begin{tabular}{ll}
      A & B\\
      \includegraphics{figure3a}&
      \includegraphics{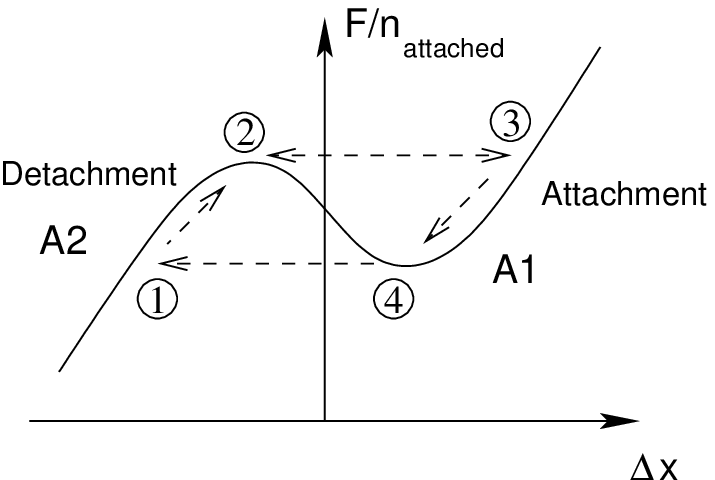}\\
      C&\\
      \multicolumn{2}{l}{\includegraphics{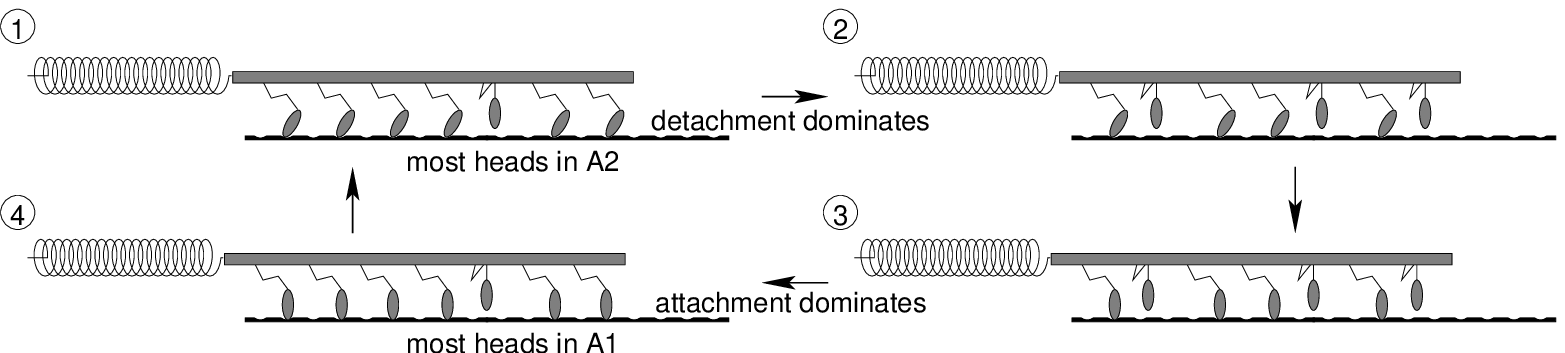}}\\
    \end{tabular}
  \end{center}
  \caption{A) A group of motors pulling against a spring. 
    The curve shows the spring extension $y$ as a function of time $t$.  When
    the motors reach stall the average velocity is zero, but fast oscillations
    occur on the length scale of the power-stroke. The observed oscillations
    are characteristic of the case where the criterion in
    Eq.~(\ref{eq:neg_slope}) is fulfilled. The simulation involved 300 motors
    pulling on an elastic element with spring constant $K_{\rm ext}=10\,{\rm
      pN/nm}$, with other parameters equal those of the dashed line in
    Fig.~\ref{fig_t2}. B,C) The mechanism of oscillations: (1) When the
    majority of bound motors are in state A2, the average detachment rate is
    higher than the attachment rate, which leads to a fall in the number of
    attached motors (2).  The force per bound motor, shown in (B), therefore
    increases. When the upper limit of the hysteresis is reached, the system
    abruptly jumps to the other fixed point in which most bound motors are in
    the state A1 (3). Because the detachment rate is low in the state A1, the
    number of bound motors increases again (4) whereby the average force per
    bound motor drops. This eventually leads to a transition back to the fixed
    point in which most motors are in state A2 (1), at which point the
    oscillation cycle repeats. }
  \label{fig_osci}
\end{figure*}

For a single pair of filaments, true isometric conditions imply that both
filaments are held fixed. However, experimental conditions are usually
near-isometric, in that the average velocity is zero, but the filaments still
have the freedom to move.  This situation occurs, for example, if both
filaments are held at their ends by a flexible spring (in which case the motors
cause the filaments to slide until the stall force is reached), or if the
filaments suffer a constant load which is precisely adjusted to prevent net
sliding. In this case the conditions are actually isotonic (or nearly so) and
the instability discussed in the previous section implies that there is no
steady state of the system with zero velocity when Eq.~(\ref{eq:neg_slope}) is
satisfied. Instead, as shown in Fig.~\ref{fig_osci}, the pair of filaments
\textit{oscillates} in near-isometric conditions. The mechanism that generates
the oscillations is the following. An ensemble of bound motors pulling against
a constant load has two stable configurations: one with a majority of motors in
state A2, the other with the majority in state A1.  Because the detachment of
motors is faster from state A2 than from state A1, the total number of bound
motors decreases when the system is in the first configuration. As a result,
the load per motor increases until the remaining motors cannot support the load
any more, whereupon the system flips into the second configuration.
Subsequently, the number of bound motors starts to grow again, and when the
load per motor falls below a critical level the system flips back to the first
configuration. Repetition of the cycle gives rise to an oscillation whose
asymmetry reflects the differences in attachment and detachment rates.

Could the isometric transient be modified if it is measured in near-isometric
conditions, instead of exactly isometric conditions? Fig.\ \ref{fig_t2} shows
the average $T_2$ response in the situation where the filaments are held at the
ends by an external spring. In the non-oscillating case ($\epsilon_1=1.33$,
dashed line) the difference between this $T_2$ curve and the one in the
strictly isometric state is only quantitative.  The most important result is
that the range of negative slope in $T_2$ remains.  On the other hand, in the
oscillating case ($\epsilon_1=0.8$, dot-dashed line) the negative slope is
flattened out because of the broader distribution of strains on cross-bridges.

\subsection{Effect of filament elasticity}
\label{sec:elasticity}

\begin{figure}
  \begin{center}
    \includegraphics{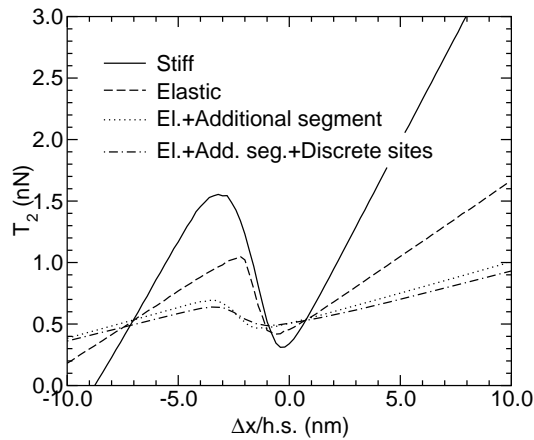}
  \end{center}
  \caption{$T_2$ curves for a stiff (solid line) and an elastic (dashed line)
    filament.  The simulations were performed with a group of $N=150$ myosin
    heads, attached to a filament of a total length $1000\,{\rm nm}$ and
    elasticity $\gamma=44000\,{\rm pN}$.  The dotted line was obtained when an
    additional segment of length $500\,{\rm nm}$ of free actin between the
    clamp and the active part was taken into account.  The dot-dashed line
    shows the same situation, but additionally taking into account discrete
    binding sites on actin. In the latter case the binding rate $k_{\rm
      bind}=100\,{\rm s}^{-1}$ was adjusted to give a realistic fraction of
    attached heads in the isometric state, therefore $\gamma_1=0.8$ and
    $\gamma_2=0.125$. All these effects together are still not able to
    completely cancel the negative slope.}
  \label{fig_t2_elastic}
\end{figure}

The compliance of the thick and thin filaments is also expected to affect the
$T_2$ curves. We model it by introducing a linear elasticity in the backbone
connecting the heads (which is, to first order, equivalent to an elasticity in
the track the motors are running on) \citep{vilfan98}.  With $\gamma$ being the
linear modulus of the filaments, which has the measured value
$\gamma=44000\,{\rm pN}$ \citep{kojima94}, the spring constant of a filament
segment of length $L$ is $K_f=L^{-1} \gamma$.  In \citep{vilfan98} a linear
two-state model was used to study the effect of filament elasticity on
force-velocity relations, showing that they reduce the isometric force if the
compliance of a filament segment between two bound motors becomes comparable to
the motor spring constant $K$.  Thin filaments are just stiff enough to prevent
significant losses due to this effect.  However, the filament compliance can
have a significantly bigger effect on the transient response. The result of a
simulation is shown in Fig.~\ref{fig_t2_elastic}. Although the filament
compliance almost halves the total stiffness of the system, it leaves the major
part of the interval of negative slope unaffected.

\subsection{Effect of discrete binding sites}

\begin{figure}
 \begin{center}
   \includegraphics{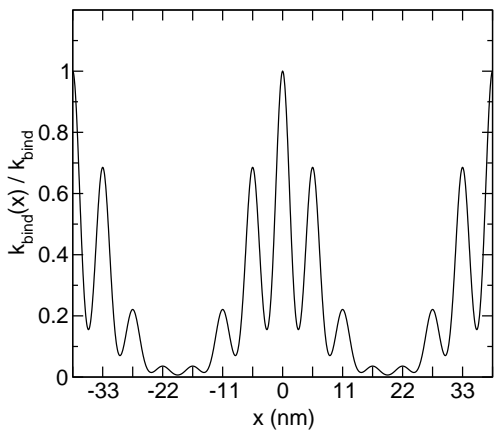}
\end{center}  
  \caption{The binding rate $k_{\rm bind}(x)/k_{\rm bind}$ of a myosin head as
    a function of its position along the actin filament.  The longitudinal
    spring constant was $K=2.5\,{\rm pN/nm}$ and the lateral spring constant
    was $K_A=15\,{\rm pNnm}$.}
  \label{fig_binding_rate}
\end{figure}

So far we have assumed that a myosin head can bind anywhere on the actin
filament with equal probability.  However, in reality the binding sites on
actin are $c=5.5\,{\rm nm}$ apart.  In addition, the actin filament has a
helical structure with a half-pitch of about $38\,{\rm nm}$ (7 monomers).  We
therefore propose that the binding rate to a particular site is proportional to
the Boltzmann factor determined by the spring distortion energy, composed of a
longitudinal and a lateral component
\begin{equation}
  \label{eq_binding_rate}
  k_{\rm bind}(\xi,\phi) = k_{\rm bind}
  \exp\left[-\frac{K\xi^2+K_A\phi^2}{ 2 k_BT}\right]
\end{equation}
with $\xi=x- (i+7j)c$ and $\phi=\pi \frac i 7$, where $j$ denotes the repeat on
the helix and $i$ the consecutive number of the actin monomer.  The total
binding rate at a given position $x$ then reads
\begin{equation}
  k_{\rm bind}(x)=k_{\rm bind}  \sum_{i,j}\exp\left[ -\frac{K(x-(i+7j)c)^2+K_A(\pi i
  /7)^2}{ 2 k_BT} \right] \;.
\end{equation}
These curves have been measured experimentally using S1 myosin heads by
\citet{Steffen.Sleep2001} and the data were fitted with $K_A=15\,{\rm pN/nm}$,
the value which we use here. The position-dependence of the binding rate for
these parameters is shown in Fig.~\ref{fig_binding_rate}.  A similar
distribution has also measured for myosin V \citep{Veigel.Molloy2002}.
 
The dot-dashed line in Fig.~\ref{fig_t2_elastic} shows the $T_2$ response when
taking into account discrete binding sites and the elasticity of actin
filaments.  As noted by \citet{Huxley.Tideswell1996}, both the filament
compliance and the discrete binding sites contribute to the flattening of the
$T_2$ curve.  However, our simulation indicates that their combined effect is
not sufficient to cancel the negative slope with the parameters used here.  We
therefore conclude that the absence of negative slope in the measured $T_2$
curves cannot be explained in terms of a single-filament model, but requires
taking into account the action of serially connected sarcomeres.

\section{Motors in muscle sarcomeres}
\subsection{Redistribution of filaments}

\begin{figure}
   \begin{center}
     \includegraphics[width=8cm]{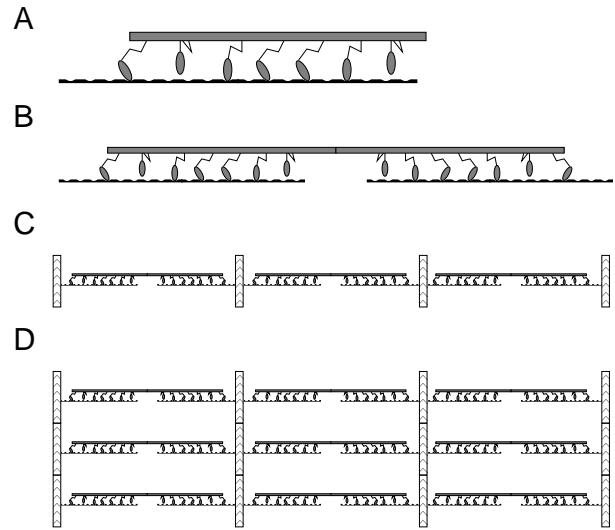}
  \end{center}
  \caption{A) A single actin filament interacting with a single myosin
    filament. B) A pair of actin filaments, interacting with a myosin filament.
    C) $n=3$ pairs of actin filaments, connected in series. D) A muscle-like
    structure with $n=3$ sarcomeres connected in series and each sarcomere
    containing $p=3$ thick filaments, connected in parallel via the Z-discs.}
  \label{fig_sarcomere}
\end{figure}

\begin{figure}
  \begin{center}
    \includegraphics{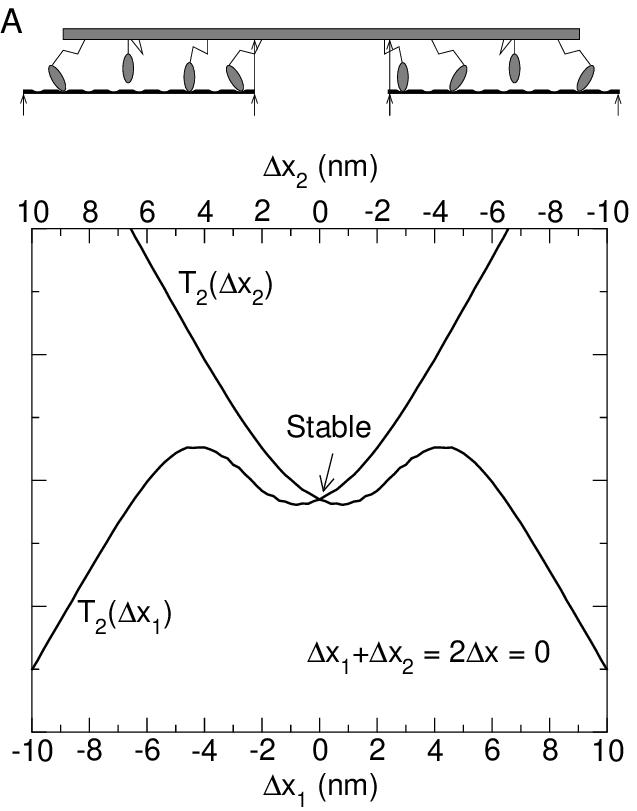} \includegraphics{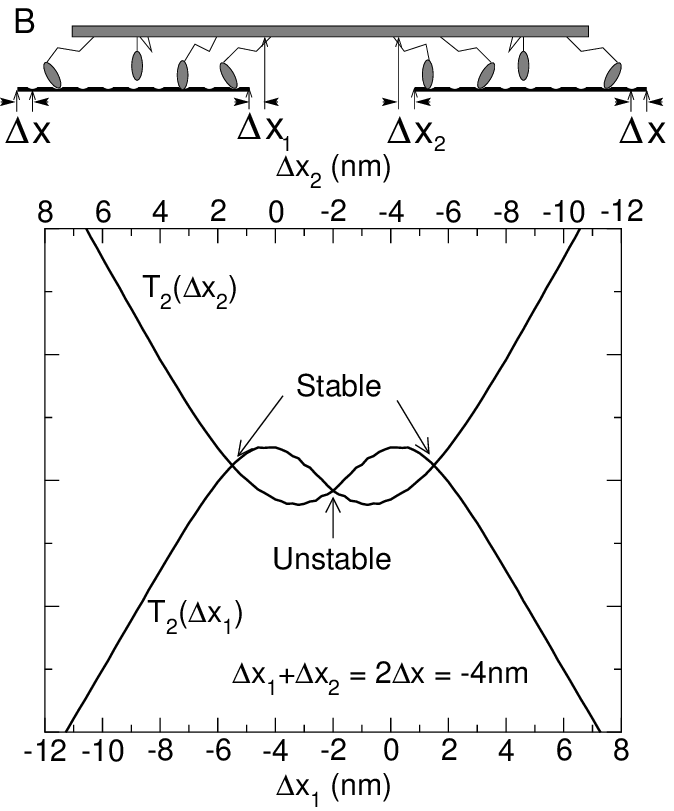}
  \end{center}
  \caption{The force-displacement relations of the left and the right half of a
    myosin filament before (A) and after (B) a length step.  The sum of their
    displacements must equal the total sarcomere stretch, $\Delta x_1 +\Delta
    x_2=2\Delta x$. This condition is imposed by plotting the
    force-displacement-relation of one set of motors as a function of $\Delta
    x_1$ ($x$-axis) and that of the other set as a function of $\Delta x_2=2
    \Delta x - \Delta x_1$ (upper $x$-axis).  The forces on the myosin filament
    are in equilibrium if the forces produced on both sides ($T_2(\Delta x_1)$
    and $T_2(\Delta x_2)$) are equal, which is given by the intersection of
    both curves.  The equilibrium is stable if the first curve crosses the
    second from below and unstable if it crosses it from above. If there are
    three stationary points (B), the central one is always unstable and those
    to either side stable.  The force-displacement relation of the two-filament
    unit is given by the force at the stable intersection as a function of the
    mean displacement $\Delta x$.}
  \label{fig_crossings}
\end{figure}

The situation changes essentially if we take into account that the experiments
are performed on whole muscle fibers, where several hundreds of sarcomeres are
connected in series, as shown in Fig.~\ref{fig_sarcomere}d.  The tension in
each half-sarcomere must be the same, but the extension of each half-sarcomere
can differ, provided that their sum equals the total stretch of the muscle
fiber. To see what happens in this situation, it is instructive first to look
at a minimum unit consisting of two filament pairs, joined back-to-back as
shown in Fig.~\ref{fig_sarcomere}b. This unit represents two thin filaments and
a thick filament, the basic building block of a sarcomere. A sudden stretch of
the unit will first displace both halves equally, according to the elasticity
of cross-bridges.  But in the next phase, when the distribution of states A1
and A2 equilibrates, the position of the thick filament can become unstable.
This means that a slight fluctuation in one direction will cause the motors
pulling the thick filament in that direction to increase their force and those
on the other side to decrease it, and the thick filament will jump sideways to
one of the stable points.  A schematic example is shown in
Fig.~\ref{fig_crossings}.

\begin{figure}
   \begin{center}
     \includegraphics{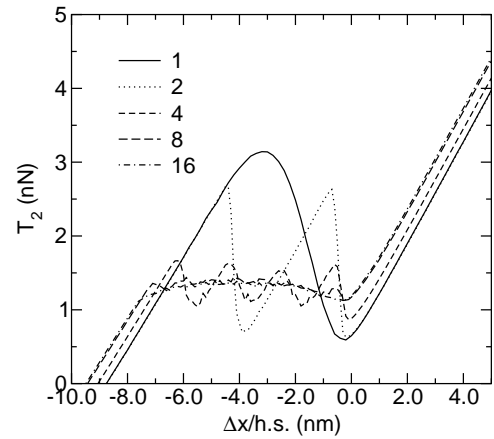}
  \end{center}
  \caption{The $T_2$ response of a single actin-myosin filament pair, compared with 2, 4, 8, and
    16 filament pairs in series.  Each myosin filament contains 300 heads.}
  \label{fig_t2_sarcomere}
\end{figure}

\begin{figure}
 \begin{center}
   \includegraphics{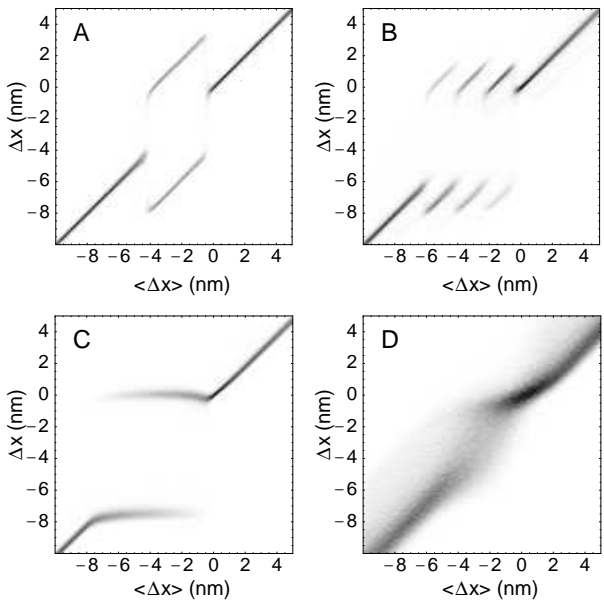}
\end{center}
\caption{Plot showing the probability density of transient stretches of
  individual half-sarcomeres (y-axis) against the total stretch per
  half-sarcomere. The three plots were obtained for (A) 2, (B) 4 and (C) 32
  filament pairs in series, each containing $N=300$ motors.  (D) shows a
  simulation with 32 filament pairs and $\epsilon_1=0.8$, in which spontaneous
  oscillations of individual filaments take place, thereby blurring the
  individual shifts.}
  \label{fig_t2_density}
\end{figure}

The result of a simulation using different numbers of filament pairs, connected
in series (the situation illustrated in Fig.~\ref{fig_sarcomere}c), is shown in
Fig.~\ref{fig_t2_sarcomere}.  The interval of negative slope is increasingly
flattened out as the number of pairs increases.  Apart from the cancellation of
the hysteresis, a shift in the curves is also observable, corresponding to the
transition between the strictly isometric conditions and those where a filament
is held under isotonic conditions before the stretch, cf.\ Fig.~\ref{fig_t2}.
The effect of redistribution can best be seen if one plots the distribution of
stretches of individual half-sarcomeres against the total stretch per
half-sarcomere (Fig.~\ref{fig_t2_density}).  For large stretch amplitudes the
distribution is concentrated on the diagonal, in agreement with the expectation
that all filaments experience the same stretch.  For two filament pairs in
series, the distribution splits in the intermediate range.  The two peaks
represent two stretches for which the single-filament $T_2$ curves produce the
same force.  Their position is determined by the condition that the average
stretch must equal the value on the $x$-axis. With four filament pairs in
series, a similar situation occurs, but with three different regimes.  First
three out of four pairs get displaced in one direction while the fourth is
displaced in the other direction.  Then this ratio changes to two against two
and finally to one against three, before the regime with all filaments being
subject to the same displacement settles in.  With higher numbers of filaments
in series (Fig.~\ref{fig_t2_density}c shows the situation for 32) this pattern
becomes increasingly continuous.  In the regime where the negative slope
stretches over the isometric point, the individual filaments perform small
oscillations (Fig.~\ref{fig_osci}) while the whole muscle is held at a constant
length. These oscillations blur the distribution further
(Fig.~\ref{fig_t2_density}d).

\begin{figure}
 \begin{center}
   \includegraphics{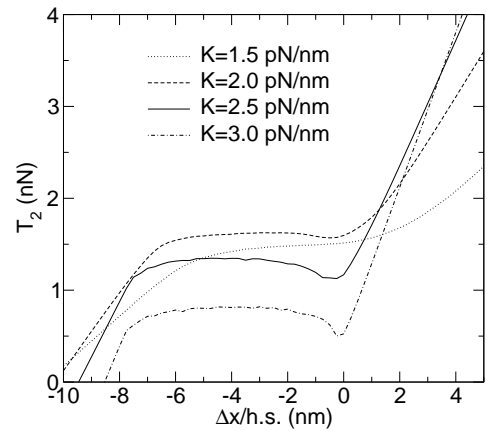}
\end{center}
\caption{$T_2$ curves with filament redistribution. The simulations were
  performed with $n=16$ filament pairs in series, each containing $N=300$
  myosin motors. Different curves show data for different values of the
  cross-bridge elasticity $K$, corresponding to values of $\epsilon_1$ between
  $0.8$ and $1.6$, and $\epsilon_0$ between $11.6$ and $23.2$.}
\label{fig_t2vsdeltag}
\end{figure}

$T_2$ curves for four different values of $K$ are shown in
Fig.~\ref{fig_t2vsdeltag}.

\subsection{Isometric state}
\label{sec:isometric}

It is instructive to take a closer look at the parameters of the globally
isometric state (by globally we mean that the total length of a fiber composed
of many sarcomeres is constant, rather than the positions of individual
filaments) in the two cases: one where the individual filaments oscillate and
one where their positions just fluctuate.  All data were determined in a
situation where a large number (32 in the simulation) of filament pairs were
connected in series and the total length held constant (situation (c) in
Fig.~\ref{fig_sarcomere}). One expects the oscillating filaments to consume
more ATP than the non-oscillating filaments.  The results are summarized in
Table \ref{tab:isometric}. An exact comparison with experimental data is
difficult, because the fraction of attached heads in the isometric state is not
known reliably, and also because the experiments were carried out at different
temperatures and with different types of myosin.  For example, the value
measured by \citet{Barsotti.Ferenczi1988}, $0.4$ ATP molecules per second per
head (with $25\%$ of the heads attached this gives $1.6\,{\rm s}^{-1}$ per
attached head), is compatible with the first scenario.  However, other
measurements give higher values \citep{Ebus.Stienen1996}, more consistent with
the second case.  We therefore conclude that from the experimental values of
the ATPase rate in the isometric state neither scenario can be excluded.

\begin{table}
  \caption{The parameters of the globally isometric state in the
    non-oscillating and the oscillating case.}
  \label{tab:isometric}
  \begin{center}
    \begin{tabular}{lcc}
      \hline
      &Non-oscillating&Oscillating\\
      Elastic constant&$K=2.5\,{\rm pN/nm}$ &$K=1.5\,{\rm pN/nm}$\\
      $\epsilon_0$ & $19.3$ & $11.6$ \\
      $\epsilon_1$ & $1.33$ & $0.8$ \\
      $\epsilon_2$ & $1.6$ & $0.95$ \\
      \hline
      Fraction heads attached & 92\% & 68\% \\
      $\frac{\text{Heads in state A2}}{\text{Attached heads}}$ & 3.5\% & 32\% \\
      $\frac{\text{Force}}{\text{attached head}}$ & $4.8\,{\rm pN}$ & $7.4\,{\rm pN}$\\
      $\frac{\text{ATPase}}{\text{attached head}}$ & $1.7\,{\rm s}^{-1}$ & $17.4\,{\rm s}^{-1}$\\
      \hline
    \end{tabular}
  \end{center}
\end{table}

At high values of $\epsilon_1$, the model shows an interesting feature. In the
isometric state the number of cross-bridges in the state A2 can be very low,
although the generated force per attached cross-bridge is as high as $5\,{\rm
  pN}$, in agreement with single-filament measurements
\citep{Kawai.Ishiwata2000}. This allows the muscle to support a load under
isometric conditions with little ATP consumption \citep{duke2000}.  It might at
first sound paradoxical that most of the isometric force is generated by the
pre-power-stroke state A1. This is possible because the myosin heads bind with
a stochastic distribution of strains.  Because those with a negative strain are
more likely to undergo the power-stroke and then detach, the remaining ensemble
produces a positive force.  In this aspect, the power-stroke actually serves to
eliminate negatively strained cross-bridges rather than direct generation of
positive force.  This notion is in agreement with recent experiments, which
have shown that high phosphate concentration does reduce the isometric force
significantly \citep{Cooke.Pate1985}, but it does not have a visible effect on
the conformation of the myosin heads or even their catalytic domains
\citep{Baker.Thomas1999}. But let us stress again that this holds only for the
isometric state.  The force in a contracting muscle originates mainly from the
state A2.

\subsection{Isotonic response}
\label{sec:isotonic}

Another important class of experiments which provides information on the
actomyosin interation involves the isotonic transient. Here the applied force
is initially set at the value of the stalling load $T_0$, so that the fiber is
prevented from contracting. The force is then suddenly changed and held
constant at a different value, while the length of the fiber is recorded. Some
early experiments showed that following a small step change of load, damped
oscillations were imposed on the steady contraction or extension of the fiber
\citep{podolsky60,Granzier90}. Such oscillations are particularly clear in
recent experiments on single muscle fibers \citep{Edman.Curtin2001}.

One possible cause of this oscillatory response has previously been suggested
on the basis of the stochastic model of the actomyosin interaction used in this
article. When $\epsilon_1 > 1$, the chemical cycles of myosin motors on the
same filament can become synchronized at loads close to the stalling force
\citep{duke99,duke2000}. A pair of filaments then slides in a step-wise fashion
under isotonic conditions.  But during steady shortening, the motors on
different filaments within the same muscle fiber operate out of phase, so that
there is no macroscopic manifestation of the steps.  However, an abrupt change
in the load can cause the synchronization of a large fraction of the bound
motors, whereupon the steps do become observable \citep{duke99}.  Because the
correlation of the motors soon decays, the macroscopic steps fade and a damped
oscillation is seen.

\begin{figure}
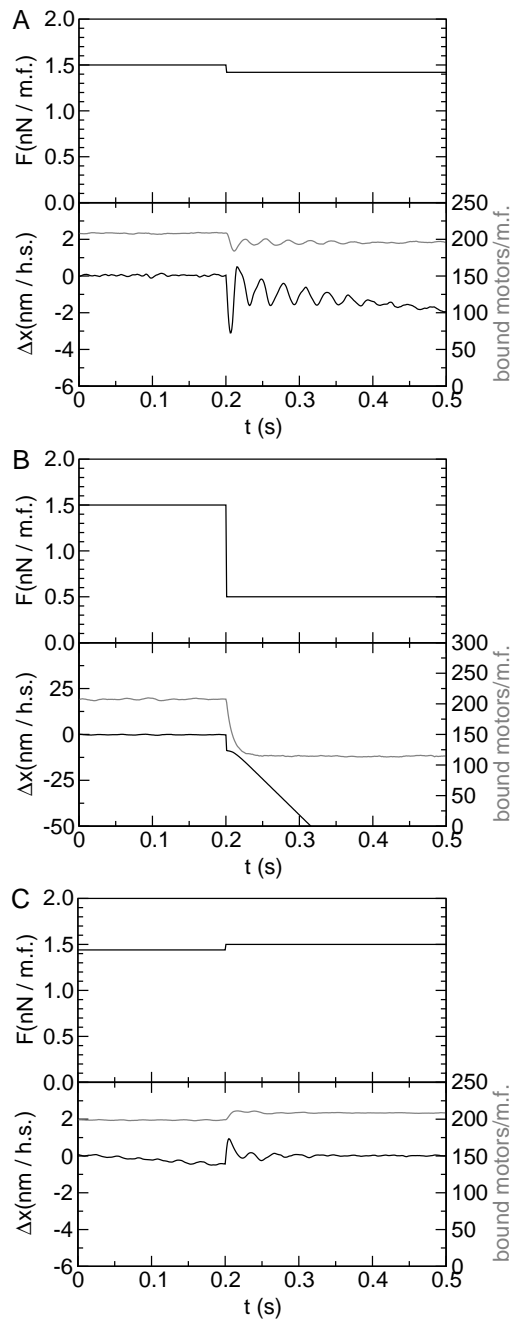

\begin{center}
  \includegraphics{figure11a}\\
  \includegraphics{figure11b}\\
  \includegraphics{figure11c}
\end{center}
\caption{The isotonic transient response of a sarcomeric structure to: (A) a small
  drop, (B) a large drop, and (C) a rise in the load, when the initial value of
  the load is chosen precisely to stall the contraction.  The upper graph shows
  the force per myosin filament as a function of time.  The lower graph
  displays the length change per half-sarcomere and the number of attached
  myosin heads per myosin filament, which is a measure of the fiber stiffness.
  The data were computed with $N=300$, $\epsilon_1=0.8$ and with 50 myosin
  filaments acting in parallel.  The results were averaged over 125 events
  (which has the same effect as simulating that number of sarcomeres in
  series). A small drop in the load synchronizes the cross-bridges and
  therefore causes observable macroscopic oscillations (A). The oscillations
  are less pronounced, but still visible following a small increase in the load
  (C).}
  \label{fig_forcestep}
\end{figure}

A much stronger oscillatory response is seen in the regime where individual
pairs of filaments perform oscillations in the near-isometric state, as
described in Sect.~\ref{sec_osci}. The synchronized oscillations can then be
very pronounced following a small decrease in the load, as shown in
Fig.~\ref{fig_forcestep}a.  On the other hand, no damped oscillations are
observed after a larger drop in the load, e.g. to $T_0/3$
(Fig.~\ref{fig_forcestep}b), because the individual filament pairs immediately
move out of the hysteretic regime.  These properties are in agreement with
recent experiments on the isotonic response of single muscle fibers carried out
by \citet{Edman.Curtin2001}.

In steady isotonic conditions, another kind of instability can arise due to a
hysteresis in the force-velocity relationship. The possibility of such an
instability was first discussed in the context of a two-state ratchet model
\citep{prost95} and subsequently in a kinetic cross-bridge model with a
strain-dependent detachment rate \citep{vilfan99}. With the parameters we use
this hysteresis, though existent, covers a rather small range of velocities and
only results in a small inflexion in the force-velocity curve
\citep{duke99,duke2000}.

\section{Discussion}
\label{sec:discussion}

In the past, the isometric transient model of muscle has been modelled by
considering the dynamics of a single pair of filaments. A significant problem
with this approach is that, with a general choice of parameters in a
cross-bridge model, the $T2$ curve is not flat \citep{hill74} --- it typically
has either a positive or a negative slope for limitingly small step
displacements. Thus, in order to reproduce the experimental $T2$ curve of
muscle, some of the parameters have needed to be finely adjusted
\citep{huxley71,eisenberg80}, and limitations have been imposed on the size $d$
of the power-stroke and the rigidity $K$ of the cross-bridges
\citep{Huxley.Tideswell1996}.  In this article we have investigated the
isometric transient response using a stochastic model of the actomyosin cycle.
When values of the parameters $d$ and $K$ are chosen to explain other
characteristics of muscle, like the force-velocity relation and the efficiency
of energy transduction, the $T_2$ curve of a single pair of filaments displays
a region of negative slope, which cannot be eliminated by factors including
filament compliance and the discrete nature of the binding sites on the actin
filament.  However, we argue that the symmetric structure of sarcomeres must be
taken into consideration when computing the isometric transient response of
muscle. Following a step change of length of a muscle fiber, a redistribution
of sarcomere lengths occurs within the fiber.  Some half-sarcomeres contract,
while others extend. This redistribution always eliminates the negative slope,
leading to a flat $T_2$ curve. It is a generic feature of unstable elements
connected in series, and does not require any special values of model
parameters.

Our model shows that there are two different regimes of the microscopic
dynamics in near-isometric conditions. For a range of values of the parameter
$\epsilon_1$ close to unity, the isometric point falls in the interval where
the slope of the $T_2$ curve is negative.  In this case individual filament
pairs oscillate with small amplitude. In the other regime, where the $T_2$
curve has positive slope at the isometric point, the individual filaments are
stationary, apart from stochastic fluctuations. The macroscopic manifestations
of these two regimes differ in few respects. Because the oscillations of
different filament pairs have different phases, oscillatory motion is not
normally observable on the scale of a whole muscle fiber in steady conditions.
However, a sudden change of load can synchronize the oscillations and thereby
make them visible. The existence of damped oscillations in the isotonic
transient response of single muscle fibers \citep{Edman.Curtin2001} therefore
argues in favor of the oscillating regime. We note, however, that damped
oscillations can also be a manifestation of step-wise shortening
\citep{duke99}, which can exist in both regimes. Because the efficient
transduction of energy demands $\epsilon_1 \approx 1$, which is close to value
of this parameter at the boundary of the two regimes, it is possible that both
regimes exist depending on conditions such as the myosin isoform, phosphate
concentration, pH, ionic strength and temperature. Indeed, measurements by
\citet{Edman.Curtin2001} show a dependence of the oscillation decay on the
solution pH and on muscle fatigue.  Further experiments, in which conditions
are systematically varied, could shed more light on the mechanism of
oscillation.

As a final remark, we emphasize that according to our model, the $T_2$ curve of
an individual pair of filaments differs from that of a muscle fiber. Recently,
assays have been developed to measure the force-velocity relation of a single
filament within a half-sarcomere \citep{Kawai.Ishiwata2000}. In order to test
our predictions, it would worthwhile to develop such high-precision techniques
to measure the transient response of a single filament.

\section*{Acknowledgment}
A.V. would like to acknowledge support from the European Union through a Marie
Curie Fellowship (No.~HPMFCT-2000-00522) and from the Slovenian Office of
Science (Grant No.~Z1-4509-0106-02).  T.D. acknowledges support from the Royal
Society.


\begin{thebibliography}{}

\bibitem[Baker et~al., 1999]{Baker.Thomas1999}
Baker, J.~E., L.~E. LaConte, I.~{Brust-Mascher}, and D.~D. Thomas. 1999{\em{}}.
\newblock {Mechanochemical coupling in spin-labeled, active, isometric muscle}.
\newblock {\em Biophys.~J.} 77:2657-2664.

\bibitem[Barclay, 1998]{Barclay1998}
Barclay, C.~J. 1998{\em{}}.
\newblock Estimation of cross-bridge stiffness from maximum thermodynamic
  efficiency.
\newblock {\em J.~Muscle\ Res.\ Cell Motil.} 19:855-864.

\bibitem[Barsotti and Ferenczi, 1988]{Barsotti.Ferenczi1988}
Barsotti, R.~J., and M.~A. Ferenczi. 1988{\em{}}.
\newblock Kinetics of {ATP} hydrolysis and tension production in skinned
  cardiac muscle of the guinea pig.
\newblock {\em J.~Biol.~Chem.} 263:16750-16756.

\bibitem[Brenner, 1991]{Brenner1991}
Brenner, B. 1991{\em{}}.
\newblock Rapid dissociation and reassociation of actomyosin cross-bridges
  during force generation: a newly observed facet of cross-bridge action in
  muscle.
\newblock {\em Proc.\ Natl.\ Acad.\ Sci.\ USA} 88:10490-10494.

\bibitem[Brenner et~al., 1995]{brenner95}
Brenner, B., J.~M. Chalovich, and L.~C. Yu. 1995{\em{}}.
\newblock Distinct molecular processes associated with isometric force
  generation and rapid tension recovery after quick release.
\newblock {\em Biophys.~J.} 68:106s-111s.

\bibitem[Chen and Brenner, 1993]{chen93}
Chen, Y., and B.~Brenner. 1993{\em{}}.
\newblock On the regeneration of the actin-myosin power stroke in contracting
  muscle.
\newblock {\em Proc.\ Natl.\ Acad.\ Sci.\ USA} 90:5148.

\bibitem[Cooke and Pate, 1985]{Cooke.Pate1985}
Cooke, R., and E.~Pate. 1985{\em{}}.
\newblock {The effects of ADP and phosphate on the contraction of muscle
  fibers}.
\newblock {\em Biophys.~J.} 48:789-798.

\bibitem[Duke, 1999]{duke99}
Duke, T. A.~J. 1999{\em{}}.
\newblock Molecular model of muscle contraction.
\newblock {\em Proc.\ Natl.\ Acad.\ Sci.\ USA} 96:2770-2775.

\bibitem[Duke, 2000]{duke2000}
Duke, T. 2000{\em{}}.
\newblock Cooperativity of myosin molecules through strain-dependent chemistry.
\newblock {\em Philos.\ Trans.\ R.\ Soc.\ Lond.\ B Biol.\ Sci.} 355:529-538.

\bibitem[Ebus and Stienen, 1996]{Ebus.Stienen1996}
Ebus, J.~P., and G.~J. Stienen. 1996{\em{}}.
\newblock {ATP}ase activity and force production in skinned rat cardiac muscle
  under isometric and dynamic conditions.
\newblock {\em J.~Mol.~Cell.~Cardiol.} 28:1747-1757.

\bibitem[Edman and Curtin, 2001]{Edman.Curtin2001}
Edman, K. A.~P., and N.~A. Curtin. 2001{\em{}}.
\newblock Synchronous oscillations of length and stiffness during loaded
  shortening of frog muscle fibres.
\newblock {\em J. Physiol.} 534:553-563.

\bibitem[Eisenberg et~al., 1980]{eisenberg80}
Eisenberg, E., T.~L. Hill, and Y.~Chen. 1980{\em{}}.
\newblock Cross-bridge model for muscle contraction.
\newblock {\em Biophys.~J.} 29:195-227.

\bibitem[Ford et~al., 1977]{ford77}
Ford, L.~E., A.~F. Huxley, and R.~M. Simmons. 1977{\em{}}.
\newblock Tension responses to sudden length change in stimulated frog muscle
  fibres near slack length.
\newblock {\em J.~Physiol.} 269:441-515.

\bibitem[Ford et~al., 1985]{Ford.Simmons1985}
Ford, L.~E., A.~F. Huxley, and R.~M. Simmons. 1985{\em{}}.
\newblock Tension transients during steady shortening of frog muscle fibres.
\newblock {\em J. Physiol.} 361:131-150.

\bibitem[Ford et~al., 1986]{Ford.Simmons1986}
Ford, L.~E., A.~F. Huxley, and R.~M. Simmons. 1986{\em{}}.
\newblock Tension transients during the rise of tetanic tension in frog muscle
  fibres.
\newblock {\em J. Physiol.} 372:595-609.

\bibitem[Granzier et~al., 1990]{Granzier90}
Granzier, H.~L., A.~Mattiazzi, and G.~H. Pollack. 1990{\em{}}.
\newblock {Sarcomere dynamics during isotonic velocity transients in single
  frog muscle fibers}.
\newblock {\em Am.\ J. Physiol} 259:C266-278.

\bibitem[Hill, 1939]{hill38}
Hill, A.~V. 1939{\em{}}.
\newblock The heat of shortening and dynamic constants of muscle.
\newblock {\em Proc. R. Soc. London Ser. B} 126:136-195.

\bibitem[Hill, 1974]{hill74}
Hill, T.~L. 1974{\em{}}.
\newblock Theoretical formalism for the sliding filament model of contraction
  of striated muscle. {P}art {I}.
\newblock {\em Prog. Biophys. Mol. Biol.} 28:267-340.

\bibitem[Howard, 2001]{Howard_book}
Howard, J. 2001{\em{}}.
\newblock {\em Mechanics of Motor Proteins and the Cytoskeleton}.
\newblock Sinauer, Sunderland, MA.

\bibitem[Huxley, 1957]{huxley57}
Huxley, A.~F. 1957{\em{}}.
\newblock Muscle structure and theories of contraction.
\newblock {\em Prog. Biophys. Biophys. Chem.} 7:255-318.

\bibitem[Huxley and Simmons, 1971]{huxley71}
Huxley, A.~F., and R.~M. Simmons. 1971{\em{}}.
\newblock Proposed mechanism of force generation in striated muscle.
\newblock {\em Nature} 233:533-538.

\bibitem[Huxley and Tideswell, 1996]{Huxley.Tideswell1996}
Huxley, A.~F., and S.~Tideswell. 1996{\em{}}.
\newblock {Filament compliance and tension transients in muscle}.
\newblock {\em J.~Muscle\ Res.\ Cell Motil.} 17:507-511.

\bibitem[Huxley and Tideswell, 1997]{Huxley.Tideswell1997}
Huxley, A.~F., and S.~Tideswell. 1997{\em{}}.
\newblock {Rapid regeneration of power stroke in contracting muscle by
  attachment of second myosin head}.
\newblock {\em J.~Muscle\ Res.\ Cell Motil.} 18:111-114.

\bibitem[Huxley, 2000]{Huxley2000}
Huxley, A.~F. 2000{\em{}}.
\newblock Cross-bridge action: present views, prospects, and unknowns.
\newblock {\em J.~Biomech.} 33:1189-1195.

\bibitem[Irving et~al., 2000]{Irving.Lombardi2000}
Irving, M., G.~Piazzesi, L.~Lucii, Y.~B. Sun, J.~J. Harford, I.~M. Dobbie,
  M.~A. Ferenczi, M.~Reconditi, and V.~Lombardi. 2000{\em{}}.
\newblock Conformation of the myosin motor during force generation in skeletal
  muscle.
\newblock {\em Nat.~Struct.~Biol.} 7:482-485.

\bibitem[J{\"u}licher and Prost, 1995]{prost95}
J{\"u}licher, F., and J.~Prost. 1995{\em{}}.
\newblock Cooperative molecular motors.
\newblock {\em Phys.~Rev.~Lett.} 75 (13):2618-2621.

\bibitem[Kawai et~al., 2000]{Kawai.Ishiwata2000}
Kawai, M., K.~Kawaguchi, M.~Saito, and S.~Ishiwata. 2000{\em{}}.
\newblock Temperature change does not affect force between single actin
  filaments and {HMM} from rabbit muscles.
\newblock {\em Biophys.~J.} 78:3112-3119.

\bibitem[Kojima et~al., 1994]{kojima94}
Kojima, H., A.~Ishijima, and T.~Yanagida. 1994{\em{}}.
\newblock Direct measurement of stiffness of single actin filaments with and
  without tropomyosin by in vitro nanomanipulation.
\newblock {\em Proc.\ Natl.\ Acad.\ Sci.\ USA} 91:12962-12966.

\bibitem[Kushmerick and Davies, 1969]{Kushmerick.Davies1969}
Kushmerick, M.~J., and R.~E. Davies. 1969{\em{}}.
\newblock The chemical energetics of muscle contraction. {II.} {T}he chemistry,
  efficiency and power of maximally working sartorius muscles.
\newblock {\em Proc.~Roy.~Soc.~London B} 174:315-347.

\bibitem[Piazzesi et~al., 2002{\em{a}}]{Piazzesi.Irving2002}
Piazzesi, G., M.~Reconditi, M.~Linari, L.~Lucii, Y.~B. Sun, T.~Narayanan,
  P.~Boesecke, V.~Lombardi, and M.~Irving. 2002{\em{a}}.
\newblock Mechanism of force generation by myosin heads in skeletal muscle.
\newblock {\em Nature} 415:659-662.

\bibitem[Piazzesi et~al., 2002{\em{b}}]{Piazzesi.Lombardi2002}
Piazzesi, G., L.~Lucii, and V.~Lombardi. 2002{\em{b}}.
\newblock The size and the speed of the working stroke of muscle myosin and its
  dependence on the force.
\newblock {\em J. Physiol.} 545:145-151.

\bibitem[Podolsky, 1960]{podolsky60}
Podolsky, R.~J. 1960{\em{}}.
\newblock Kinetics of muscular contraction: the approach to the steady state.
\newblock {\em Nature} 188:666.

\bibitem[Steffen et~al., 2001]{Steffen.Sleep2001}
Steffen, W., D.~Smith, R.~Simmons, and J.~Sleep. 2001{\em{}}.
\newblock Mapping the actin filament with myosin.
\newblock {\em Proc.\ Natl.\ Acad.\ Sci.\ USA} 98:14949-14954.

\bibitem[Tyska and Warshaw, 2002]{Tyska.Warshaw2002}
Tyska, M.~J., and D.~M. Warshaw. 2002{\em{}}.
\newblock The myosin power stroke.
\newblock {\em Cell~Motil.~Cytoskeleton} 51:1-15.

\bibitem[Veigel et~al., 1998]{veigl98}
Veigel, C., M.~L. Bartoo, C.~S. White, J.~S. Sparrow, and J.~E. Molloy.
  1998{\em{}}.
\newblock The stiffness of rabit skeletal acotmyosin cross-bridges determined
  with an optical tweezers transducer.
\newblock {\em Biophys.~J.} 75:1424.

\bibitem[Veigel et~al., 2002]{Veigel.Molloy2002}
Veigel, C., F.~Wang, M.~L. Bartoo, J.~R. Sellers, and J.~E. Molloy.
  2002{\em{}}.
\newblock {The gated gait of the processive molecular motor, myosin V}.
\newblock {\em Nat.~Cell~Biol.} 4:59-65.

\bibitem[Vilfan et~al., 1998]{vilfan98}
Vilfan, A., E.~Frey, and F.~Schwabl. 1998{\em{}}.
\newblock Elastically coupled molecular motors.
\newblock {\em Eur.\ Phys.\ J. B} 3:535-546.

\bibitem[Vilfan et~al., 1999]{vilfan99}
Vilfan, A., E.~Frey, and F.~Schwabl. 1999{\em{}}.
\newblock Force-velocity relations of a two-state crossbridge model for
  molecular motors.
\newblock {\em Europhys.~Lett.} 45:283-289.

\end{thebibliography}
\end{document}